\documentclass[a4paper,fleqn,usenatbib]{mnras}

\usepackage[T1]{fontenc}
\usepackage{ae,aecompl}

\usepackage{bm}

\usepackage{graphicx}	
\usepackage{amsmath}	
\usepackage{amssymb}	

\title[Self-trapping of g-mode oscillations]{Self-Trapping of G-Mode Oscillations in Relativistic Thin Disks, Revisited}

\author[Shoji Kato]{
Shoji kato $^{1}$\thanks{E-mail: kato.shoji@gmail.com}
\\
$^{1}$2-2-2 Shikanodai Nishi, Ikomashi, Nara, Japan, 630-0114}

\date{Accepted XXX. Received YYY; in original form ZZZ}

\pubyear{2016}

\begin{document}
\label{firstpage}
\pagerange{\pageref{firstpage}--\pageref{lastpage}}
\maketitle

\begin{abstract}
We examine by a perturbation method how the self-trapping of g-mode oscillations in 
geometrically thin relativistic disks is affected by uniform vertical magnetic fields.
Disks which we consider are isothermal in the vertical direction, but are truncated at a
certain height by presence of hot coronae.
We find that the characteristics of self-trapping of axisymmetric g-mode oscillations in 
non-magnetized disks is kept unchanged in magnetized disks at least till a strength of the fields,
depending on vertical thickness of disks.
These magnetic fields become stronger as the disk becomes thinner.
This result suggests that trapped g-mode oscillations still remain as one of possible candidates of
quasi-periodic oscillations observed in black-hole and neutron-star X-ray binaries in the cases where vertical magnetic fields in disks are weak. 
\end{abstract}




\section{Introduction}

In black-hole and  neutron-star X-ray binaries quasi-periodic oscillations (QPOs) are occasionally 
observed with frequencies close to the relativistic Keplerian frequencies of the innermost region of disks 
surrounding the central sources (e.g., van der Klis 2000; Remillard \&
McClintock 2006). 
One of possible origins of these quasi-periodic oscillations is disk oscillations in
the innermost region of the disks (discoseismology).

The innermost regions of relativistic disks are subject to strong gravitational field of 
central sources, and thus behaviors of disk oscillations in these relativistic regions are
different from those in Newtonian disks (see,  e.g., Kato and Fukue 1980).  
One of the differences is related to
the difference of radial distribution of (radial) epicyclic frequency $\kappa(r)$.
In relativistic disks the epicyclic frequency does not monotonically increase inwards, but 
has a maximum, $\kappa_{\rm max}$, at a certain radius, say $r_{\rm max}$, and
inside the radius it decreases inward to vanish at the radius of ISCO (innermost stable circular orbit).
This radius is roughly the inner edge of the disks.

The g-mode oscillations\footnote{
See, for example, Kato (2016) for classification of disk oscillations in geometrically
thin disks.
Terminology in classification by Kato (2016) is slightly different 
from that by Kato et al (2008). 
}
are oscillation modes whose frequencies, say $\omega$, are lower than 
$\kappa_{\rm max}$, i.e., $\omega^2<\kappa_{\rm max}^2$.
Okazaki et al. (1987) showed that in  relativistic disks 
the propagation region of the oscillations is bounded by the two radii where
$\omega^2=\kappa_{\rm max}^2$ is realized.
At these two radii the oscillations are reflected.
This is the self-trapping of the g-mode oscillations in relativistic disks.
Eigen-frequencies of  these trapped g-mode oscillations in geometrically thin disks are calculated in detail 
by Perez et al. (1997), using  the Kerr geometry.
Nowak et al. (1997) suggested that these trapped oscillations might be the origin of the 67 Hz oscillations observed in the black-hole source GRS 1915+105.

In the above studies, however, effects of magnetic fields on oscillations were  outside considerations.
Fu and Lai (2009) examined these effects and found important results.
That is, they showed that the g-mode oscillations are strongly affected if poloidal magnetic fields
are present in disks and the self-trapping of g-mode oscillations is destroyed even when the
fields are weak. 
For example, they showed that $c_{\rm A}/c_{\rm s}\geq 0.03$ 
(i.e., $c_{\rm A}^2/c_{\rm s}^2\geq 10^{-3}$) is enough to destroy the self-trapping
in the case of axisymmetric g-mode oscillations, where $c_{\rm A}$ and $c_{\rm s}$ are Alfv\'{e}n
and acoustic speeds in disks, respectively.

Their results  are very interesting in considering what kinds of disk oscillations are possible 
candidates of
quasi-periodic oscillations, but their analyses are rough and more careful examinations 
are necessary to evaluate a quantitative condition of destruction of the self-trapping of g-mode 
oscillations.
For example, they do not take into account quantitatively the effects of vertical 
structure of disks.
In geometrically thin disks, the disk density decreases apart from the equatorial 
plane, while 
the Alfv\'{e}n speed increases greatly in the vertical direction, if the strength of poloidal 
magnetic fields is nearly constant in disks.
These sharp decrease of density and sharp increase of Alfv\'{e}n speed in the vertical direction 
should be carefully taken into account in examination of wave motions. 

To avoid mathematical complication and difficulty related to the above situations,
it is worthwhile to consider disks with finite vertical thickness.
This is, however, not only for avoidance of mathematical complication. 
Consideration of such disks is important from the observational
and physical  viewpoints.
Many black-hole and neutron-star X-ray binaries show both soft and hard
components  of spectra, and record a soft-hard transition.
Usually three states are known:
high/soft state, very high/intermediate state (steep power-law state), and the low/hard state.
Remillard (2005) shows that detection of high frequency QPOs in 
black-hole X-ray binaries is correlated with power-law luminosity, and their frequencies are 
related to soft components.
This suggests that at the state where high frequency QPOs are
observed, the sources have both geometrically thin disks and coronae.

In this context, by using the above-mentioned disk geometry  (geometrically thin cool disk 
and a hot corona surrounding the disk),
we examine in this paper 
the effects of vertical magnetic fields on g-mode oscillations by a perturbation method.
The results  show that the perturbation method is applicable till a critical strength of magnetic fields 
which depends on disk thickness.
In the case where disk thickness is three times the vertical half-thickness, $H$, of disks, for example, 
the critical strength of magnetic fields where the perturbation method is applicable 
is given by $c_{{\rm A}0}^2/c_{\rm s}^2 \sim 0.14$, 
where $c_{{\rm A}0}$ and 
$c_{\rm s}$  are Alfv\'{e}n speed on the disk equator and acoustic speed, respectively.
Our results show that till these critical magnetic fields where the perturbation method is applicable, axisymmetric g-mode oscillations are still self-trapped.
This result is different from that obtained by use of a rough local approximation for
vertically extended disks by Fu and Lai (2009).  

\section{Unperturbed Disk Model and Basic Equations Describing Perturbations}

We are interested in g-mode oscillations in relativistic disks.
The general relativistic treatments of dynamical equations are, however, very complicated.
Hence, following a conventional way, we treat Newtonian equations, but adopt the
general relativistic expression for (radial) epicyclic frequency $\kappa(r)$.
 
We consider vertically isothermal, geometrically thin disks.
The disks are assumed to be subject to vertical magnetic fields.
We adopt the cylindrical coordinates ($r$, $\varphi$, $z$) whose origin is at the center of a central object 
and the $z$-axis is perpendicular to the disk plane.
Then, the magnetic fields in the unperturbed state, $\bm{B}_0$, are  
\begin{equation}
       \bm{B}_0(r)=[0, 0, B_0(r)].
\label{unifom-fields}
\end{equation}
Since the magnetic fields are purely vertical, they have no effects on the vertical structure of disks,
and the hydrostatic balance in the vertical direction in disks gives the disk density, $\rho_0$, stratified in the vertical direction as (e.g., see Kato et al. 2008)
\begin{equation}
          \rho_0(r, z)=\rho(r)_{00}\, {\rm exp}\biggr(-\frac{z^2}{2H^2}\biggr),
\label{density}
\end{equation}
where the scale height, $H(r)$, is related to the isothermal acoustic speed, $c_{\rm s}(r)$, and
vertical epicyclic frequency, $\Omega_\bot(r)$, by
\begin{equation}
       H^2(r)=\frac{c_{\rm s}^2}{\Omega_\bot^2}.
\label{scaleheight}
\end{equation}
In relativistic disks under the Schwarzschild metric, $\Omega_\bot$ is equal to the
relativistic Kepler frequency.

\subsection{Termination of disk thickness by hot corona}

The above-mentioned vertically isothermal disks can extend infinitely in the vertical direction, 
although they are derived under the assumption that the disks are thin in the vertical direction.
Here, we assume that the disks are terminated at a finite height, say $z_{\rm s}$, by presence of hot corona.
The height $z_{\rm s}$ is taken to be a parameter.
As a dimensionless parameter we adopt $\eta_{\rm s}\equiv z_{\rm s}/H$ hereafter.

\subsection{Equations describing perturbations}

We consider small-amplitude adiabatic perturbations in the above-mentioned vertically isothermal disks.
The perturbations are assumed to be isothermal in the vertical direction.

Since the magnetic fields, $\bm{B}_0$ are perturbed to $\bm{B}=(b_r, b_\varphi, B_0+b_z)$, 
we have
\begin{equation}
    {\rm curl}\ \bm{b}\times\bm{B}_0=\biggr[B_0\biggr(\frac{\partial b_r}{\partial z}-\frac{\partial b_z}{\partial r}\biggr),
                    -B_0\biggr(\frac{\partial b_z}{r\partial \varphi}-\frac{\partial b_\varphi}{\partial z}\biggr),0\biggr].
\end{equation}
Hence, the time variations of Eulerian velocity perturbations over the rotation,  
($u_r$, $u_\varphi$, $u_z$), are expressed by use of the $r$-, $\varphi$- and $z$-components of equation of motion, respectively, as
\begin{equation}
     \biggr(\frac{\partial}{\partial t}+\Omega\frac{\partial}{\partial\varphi}\biggr)u_r-2\Omega u_\varphi=-\frac{\partial h_1
     }{\partial r}+\frac{c_{\rm A}^2}{B_0}\biggr(\frac{\partial b_r}{\partial z}-\frac{\partial b_z}{\partial r}\biggr),
\label{ur0}
\end{equation}
\begin{equation}
     \biggr(\frac{\partial}{\partial t}+\Omega\frac{\partial}{\partial\varphi}\biggr)u_\varphi
            +\frac{\kappa^2}{2\Omega} u_r  =-\frac{\partial h_1}{r\partial\varphi}
            -\frac{c_{\rm A}^2}{B_0}\biggr(\frac{\partial b_z}{r\partial \varphi}-\frac{\partial b_\varphi}{\partial z}\biggr),
\label{uvarphi}
\end{equation}
\begin{equation}
     \biggr(\frac{\partial}{\partial t}+\Omega\frac{\partial}{\partial\varphi}\biggr)u_z=-\frac{\partial h_1}{\partial z},
\label{uz}
\end{equation}
where $\kappa(r)$ is the epicyclic frequency, $\Omega(r)$ is the angular velocity of disk
rotation, $h_1$ is
\begin{equation}
           h_1=\frac{p_1}{\rho_0}=c_{\rm s}^2\frac{\rho_1}{\rho_0},
\end{equation}
and $c_{\rm A}$ is the Alfv\'{e}n speed defined by
\begin{equation}
       c_{\rm A}^2(r, z)=\frac{B_0^2}{4\pi\rho_0}=c_{{\rm A}0}^2(r){\rm exp}\biggr(\frac{z^2}{2H^2}\biggr),
\label{Alfven}
\end{equation}
$c_{{\rm A}0}$ being the Alfv\'{e}n speed on the equator. 

The time variation of magnetic fields is governed by the induction equation, which is 
\begin{equation}
      \frac{\partial b_r}{\partial t}=\frac{\partial}{\partial z}(u_rB_0),
\end{equation}
\begin{equation}
      \frac{\partial b_\varphi}{\partial t}=\frac{\partial}{\partial z}(u_\varphi B_0),
\end{equation}
\begin{equation}
      \frac{\partial b_z}{\partial t}=-\frac{\partial}{r\partial r}(ru_rB_0)-\frac{\partial}{r\partial \varphi}(u_\varphi B_0).
\end{equation}
Finally, the time variation of density is governed by the equation of continuity, which is
\begin{equation}
     \biggr(\frac{\partial}{\partial t}+\Omega\frac{\partial}{\partial \varphi}\biggr)\rho_1
        +\frac{\partial}{r\partial r}(r\rho_0u_r)+\frac{\partial}{r\partial \varphi}(\rho_0u_\varphi)
           +\frac{\partial}{\partial z}(\rho_0u_z)=0.
\label{continuity}
\end{equation}

Hereafter, we concentrate our attention only on axisymmetric oscillations, i.e., 
$\partial/\partial \varphi=0$.
This is because we are interested in g-mode oscillations in this paper, and asymmetric g-mode
oscillations are damped by corotation resonance (Kato 2003, Li et al. 2003).
In addition, the radial variations of unperturbed quantities, such as $\Omega(r)$, 
$\rho_{00}(r)$, $H(r)$, $c_{\rm s}^2(r)$
and $c_{\rm A}^2(r)$, are neglected, assuming that the radial wavelengths of 
perturbations are shorter
than the characteristic radial lengths of unperturbed quantities (local approximations
in the radial direction).
The radial variation of $\kappa(r)$ is, however, carefully taken into account, because its radial variation
is sharp near to the disk edge.
Furthermore, $\partial/\partial t$ is written as $i\omega$, since we are interested in normal mode 
oscillations whose frequency is $\omega$.

Then, combining equations (\ref{ur0}) and (\ref{uvarphi}) we have
\begin{equation}
     -(\omega^2-\kappa^2)u_r+i\omega \frac{\partial h_1}{\partial r}
     =\frac{c_{\rm A}^2}{B_0}\biggr[i\omega\biggr(\frac{\partial b_r}{\partial z}-\frac{\partial b_z}{\partial r}
         \biggr)+2\Omega\frac{\partial b_\varphi}{\partial z}\biggr].
\label{h1ur2}
\end{equation}
The righthand side of this equation shows how the relation between $h_1$ and $u_r$ is affected by
magnetic fields.
There is another relation between $u_r$ and $h_1$.
This is obtained from equations (\ref{uz}) and (\ref{continuity}) by eliminating $u_z$, which is
\begin{equation}
     \biggr(\frac{\partial^2}{\partial z^2}-\frac{z}{H^2}\frac{\partial}{\partial z}+\frac{\omega^2}{c_{\rm s}^2}\biggr)h_1-i\omega \frac{\partial u_r}{\partial r}=0.
\label{h1ur}
\end{equation}

If $u_r$ is eliminated from the lefthand side of equation (\ref{h1ur}) by using equation (\ref{h1ur2}),
we have  
\begin{eqnarray}
     \biggr(\frac{\partial^2}{\partial \eta^2}-\eta\frac{\partial}{\partial \eta}
     +\frac{\omega^2}{\Omega_\bot^2} \biggr) h_1
     +H^2\frac{\partial}{\partial r}\biggr(\frac{\omega^2}{\omega^2-\kappa^2}\frac{\partial h_1}{\partial r}\biggr)
              \nonumber   \\
     =-i\omega H^2\frac{\partial}{\partial r}
         \biggr[\frac{c_{\rm A}^2/B_0}{\omega^2-\kappa^2}
         \biggr\{i\omega\biggr(\frac{\partial b_r}{H\partial \eta}-\frac{\partial b_\varphi}{\partial r}\biggr)
           +2\Omega\frac{\partial b_\varphi}{H\partial \eta}\biggr\}\biggr],
\label{eqh1}
\end{eqnarray} 
where the vertical coordinate $z$ has been changed to dimensionless coordinate $\eta$ defined by 
$\eta=z/H$.
By using the induction equations we can write the righthand side of equation (\ref{eqh1}) in terms of 
$\bm{u}$.
Then, equation (\ref{eqh1}) is expressed as
\begin{eqnarray}
     \biggr(\frac{\partial^2}{\partial \eta^2}-\eta\frac{\partial}{\partial \eta}
     +\frac{\omega^2}{\Omega_\bot^2} \biggr) h_1
     +H^2\frac{\partial}{\partial r}\biggr(\frac{\omega^2}{\omega^2-\kappa^2}\frac{\partial h_1}{\partial r}\biggr)
              \nonumber   \\
     =-i\omega\frac{\partial}{\partial r}
         \biggr[\frac{c_{\rm A}^2}{\omega^2-\kappa^2}
         \biggr(H^2\frac{\partial^2 u_r}{\partial r^2}+\frac{\partial^2u_r}{\partial \eta^2}
           +\frac{2\Omega}{i\omega}\frac{\partial^2u_\varphi}{\partial \eta^2}\biggr)\biggr].
\label{eqh1u}
\end{eqnarray}
 
In the case of no magnetic fields, the righthand side of equation (\ref{eqh1u}) vanishes, and we
have 
\begin{equation}
\biggr(\frac{\partial^2}{\partial \eta^2}-\eta\frac{\partial}{\partial \eta}
     +\frac{\omega^2}{\Omega_\bot^2} \biggr) h_1
     +H^2\frac{\partial}{\partial r}\biggr(\frac{\omega^2}{\omega^2-\kappa^2}\frac{\partial h_1}{\partial r}\biggr)
     =0.
\label{eqh1_0}
\end{equation}
This is the well-known wave equation in disks in a simplified situation (e.g., Kato 2001,
Kato 2016).
We examine the effects of magnetic fields on the wave motions described by equation (\ref{eqh1_0})
by a perturbation method in the next section.
 
\subsection{Boundary condition}                                         
   
As mentioned before we are interested in disks which are terminated at certain height by presence 
of hot corona.
The disk thickness is assumed to be terminated at $z=z_{\rm s}$ (i.e., at $\eta=\eta_{\rm s}
\equiv z_{\rm s}/H$).
The disk thickness, $\eta_{\rm s}$,  is a parameter, and we
impose a boundary condition at  $\eta_{\rm s}$.

It will be relevant to assume that at the deformed boundary surface between disk and corona  
$\rho u_n^2+p+B_t^2/8\pi$ is continuous\footnote{
The equation of motion can be written in such a conservative form as
$$
    \frac{\partial \rho u_i}{\partial t}+\frac{\partial}{\partial r_j}
    \biggr[\rho u_iu_j+\delta_{ij}+\frac{B^2}{8\pi}\delta_{ij}-\frac{1}{4\pi}B_iB_j\biggr]=0.
$$
The normal component of this equation shows that 
$$
                \rho u_n^2+p+\frac{1}{8\pi}B_t^2
$$
must be continuous at the boundary, where the subscript $n$ and $t$ denote, respectively, the normal and
tangential components at the boundary, as is known in the field of hydromagnetic shocks.
In deriving this equation continuity of $B_{\rm t}$ has been used.
The tangential component of the above equation of motion gives that
$$
     \rho u_n\bm{u}_t-\frac{1}{4\pi}B_n\bm{B}_t
$$
is also continuous at the boundary.
This latter condition is, however, not used here.
},
where $u_n$ and $B_t$ are the normal component of fluid velocity and the tangential component of 
magnetic fields.
By the definition of the   boundary surface we have $u_n=0$ at the surface.
Furthermore, since the unperturbed magnetic fields have no horizontal components, 
the Lagrangian variation of $B_t^2$ vanishes there, i.e., $\delta B_t^2=0$, where $\delta$ represents Lagrangian variation.
Hence, the boundary condition we should adopt is continuation of $\delta p$ at the surface.
Furthermore, we assume that in the corona,  pressure perturbations resulting from wave motions in disks
are quickly smoothed out by high temperature (large acoustic speed), i.e., we assume $\delta p=0$ in the
corona.
The timescale of the smoothing in corona is shorter than the timescale of oscillations, 
because the corona temperature will be on the order of virial temperature.
Consequently, the boundary condition we adopt is
\begin{equation}
        \delta p= p_1+\bm{\xi}\cdot\nabla p_0=0 \quad {\rm at}\ \eta=\eta_{\rm s},
\label{boundary}
\end{equation}
where $\bm{\xi}$ is displacement vector associated with perturbations.
In the case of axially symmetric perturbations ($\partial/\partial \varphi=0$), equation (\ref{boundary}) is 
written as
\begin{equation}
      h_1+\xi_r\frac{1}{\rho_0}\frac{\partial p_0}{\partial r}-\Omega_\bot^2z\xi_z=0.
\label{bc2}
\end{equation}

Combination of equations
(\ref{bc2}) and (\ref{continuity}) leads the boundary condition to 
\begin{equation}
      \frac{\partial u_r}{\partial r}+\frac{\partial u_z}{\partial z}=0 \ \ {\rm at}\ \ \eta=\eta_s.
\label{bc3}
\end{equation}
This boundary condition is further reduced by using equations (\ref{continuity}) and (\ref{uz}) in a form
expressed in terms of $h_1$ alone, which is
\begin{equation}
      \biggr(-\eta\frac{\partial}{\partial \eta}+\frac{\omega^2}{\Omega_\bot^2}\biggr)h_1=0  \ \ {\rm at}\ \ \eta=\eta_{\rm s}.
\label{bc4}
\end{equation}
Our subject is thus to solve equation (\ref{eqh1u}) with boundary condition (\ref{bc4}).

\section{Perturbation Method}  

Here, we solve equation (\ref{eqh1u}) by a perturbation method.
That is, we start from the limit of no magnetic fields, and examine the effects of magnetic fields 
by a perturbation method.

\subsection{Zeroth-order solution}

In the limit of no magnetic fields, equation (\ref{eqh1u}) is reduced to equation (\ref{eqh1_0}) and
this latter equation can be solved easily by a variable separation method.
That is, by separating $h_1(r,\eta)$ as $h_1(r,\eta)=g(\eta)f(r)$ and dividing equation (\ref{eqh1_0})
by $h_1$, 
we can separate equation (\ref{eqh1_0}) into $r$- and $\eta$- dependent parts.
Then, by introducing a separation constant, $K$, we have two equations:  
\begin{equation}
     \biggr(\frac{d^2}{d\eta^2}-\eta\frac{d}{d\eta}+K \biggr)g(\eta) =0,
\label{eqg}
\end{equation}
and
\begin{equation}
      \frac{d}{dr}\biggr(\frac{\omega^2}{\omega^2-\kappa^2}\frac{d}{dr}\biggr)f(r)
       +\frac{\omega^2-K\Omega_\bot^2}{c_{\rm s}^2} f(r)=0.
\label{eqf}
\end{equation}
The separation constant, $K$,  is determined by the boundary condition (\ref{bc4}).

Equation (\ref{eqg}) has two independent solutions, which are plane-symmetric 
(even function of $\eta$) and plane-asymmetric (odd function of $\eta$).
Their formal solutions are
\begin{equation}
     g(\eta)=1-\frac{1}{2!}K\eta^2-\frac{1}{4!}K(2-K)\eta^4-\frac{1}{6!}K(2-K)(4-K)\eta^6-...
\label{geq1}
\end{equation}
and 
\begin{eqnarray}
     g(\eta)=\eta+\frac{1}{3!}(1-K)\eta^3+\frac{1}{5!}(1-K)(3-K)\eta^5     \nonumber \\  
          + \frac{1}{7!}(1-K)(3-K)(5-K)\eta^7+...
\label{geq2}
\end{eqnarray}
In the case where the disk extends infinitely in the vertical direction, i.e., $\eta_{\rm s}=\infty$,
equations (\ref{geq1}) and (\ref{geq2}) show that  the series in these equations must be terminated 
at finite terms.
Otherwise, the boundary condition that energy density of perturbations do not diverge at infinity
cannot be satisfied. 
This requires that the separation constant $K$ [which is also eigenvalue of equation (\ref{eqg})]
needs to be zero or positive integers, i.e., the set of $K$'s are $K_n$, where  $n=$ 0, 1, 2, ..
and $K_n=0$, 1, 2,...
The eigen-function corresponding to $K_n$, say  $g_n(\eta)$,  is the Hermite polynomial, i.e., 
$g_n(\eta)={\cal H}_n(\eta)$ (Okazaki et al. 1987).

We are interested in this paper g-mode oscillations which are fundamental in their behavior in the
vertical direction, i.e., $n=1$. 
Hence, $K$ and $g$ which will appear hereafter without subscript are $K_1$ and $g_1$, respectively. 

Distinct from the case of disks which extend infinitely in the vertical direction, 
we impose in this paper a boundary condition (\ref{bc4}) at  a finite height $\eta_{\rm s}$.
In this case  $K_n$ is no longer zero nor positive integers, and equations (\ref{geq1}) and 
(\ref{geq2}) are infinite series (not terminated). 
As boundary condition (\ref{bc4}) shows, $K_n$ and $g_n(\eta)$ depend on $\eta_{\rm s}$ and 
$\omega^2/\Omega_\bot^2$.
By using expressions for $g$ given by equations (\ref{geq1}) and (\ref{geq2}), we have numerically
calculated eigenvalue $K_n$ and eigenfunction $g_n(\eta)$.
The results in the case of  $n=1$ are shown in figures 1 and 2.
Figure 1 is for $K$,  and figure 2 is  for functional forms of $g(\eta)$.
Figure 1 shows that in the limit of $\eta_{\rm s}=\infty$, $K$ tends to $K=1$ as expected.
As this figure shows, deviation from $K=1$ is large when the disk is thin ($\eta_{\rm s}$ is small)
and  $\omega^2/\Omega_\bot^2$ is small.
The value of $\omega^2/\Omega_\bot^2$ is a parameter at the present stage.
It is determined after wave equation (\ref{eqf}) in the radial direction is solved
with the boundary condition.
The eigenfunction, $g(\eta)$, in figure 2 
shows that eigenfunctions are close to $g(\eta)=\eta$, and the deviation
from $g(\eta)=\eta$ occurs only near to the boundary at $\eta=\eta_{\rm s}$.

\begin{figure}
 \begin{center}
  \includegraphics[width=8cm]{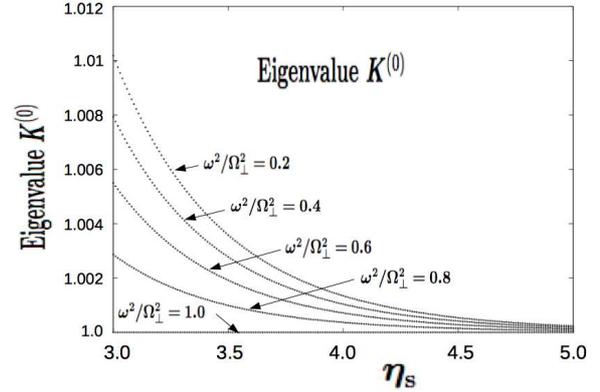} 
 \end{center}
\caption{Eigenvalues of the fundamental ($n=1$) g-mode oscillations as functions of
$\eta_{\rm s}$. 
Eigenvalues depend on frequencies of oscillations, $\omega^2$, normalized by $\Omega_\bot^2$
as shown by boundary condition (\ref{bc4}).
In the case of $\eta_{\rm s}=\infty$, eigenvalue tends to unity,  corresponding to the case of infinitely
extended isothermal disks.  
}
\label{fig:sample}
\end{figure}
\begin{figure}
 \begin{center}
  \includegraphics[width=8cm]{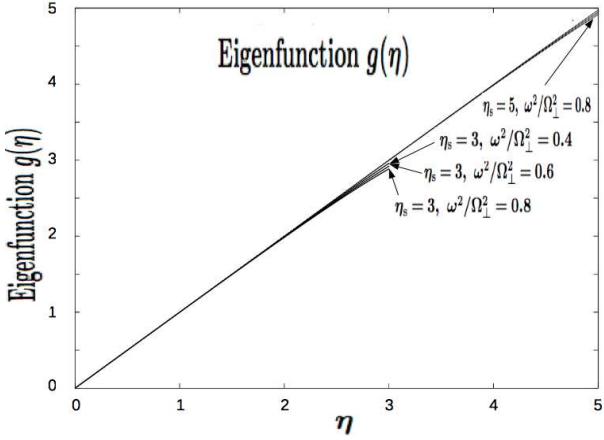} 
 \end{center}
\caption{Functional forms of eigenfunction, $g(\eta)$, of the fundamental ($n=1$) g-mode oscillations.
Their dependences on $\eta_{\rm s}$ and $\omega^2/\Omega_\bot^2$ are shown.
Two cases where disk boundary $\eta_{\rm s}$ is at 3.0 and 5.0 are shown for some values of 
$\omega^2/\Omega_\bot^2$.
It is noted that the eigenfunctions are terminated at $\eta=\eta_{\rm s}$,
but their difference from the eigenfunction for $\eta_{\rm s}=\infty$ (i.e., $g(\eta)={\cal H}_1(\eta)=\eta$),
is quite small.  
}
\label{fig:sample}
\end{figure}

\subsection{Quasi-orthogonality of zeroth-order eigenfunctions}

In the case of $\eta_{\rm s}=\infty$, the series of the eigenvalues, $K_n$, are 
$K_n=0$, 1, 2,..., and  the corresponding eigenfunctions, say $g_n$,  are orthogonal 
in the sense that 
\begin{equation}
         \int_{-\infty}^\infty{\rm exp}\biggr(-\frac{\eta^2}{2}\biggr)g_ng_m d\eta= n! (2\pi)^{1/2}\delta_{nm},
\label{orth0}
\end{equation}
because $g_n$ is the Hermite polynomial of order $n$.

The presence of orthogonality among the zeroth order solutions is helpful in applying
perturbation methods.
In the present case of $\eta_{\rm s}\not= \infty$, however, there is no such orthogonality.
In spite of this, we have quasi-orthogonality as shown below,
unless $\eta_{\rm s}$ is too small.

If the eigenfunction corresponding to $K_n$ is written as $g_n$, the zeroth-order wave equation is 
expressed in the form
\begin{equation}
     \frac{d}{d\eta}\biggr[{\rm exp}\biggr(-\frac{\eta^2}{2}\biggr)\frac{dg_n}{d\eta}\biggr]
                          +{\rm exp}\biggr(-\frac{\eta^2}{2}\biggr)K_ng_n=0.
\label{orth11}
\end{equation}
Multiplying both sides of equation (\ref{orth11}) by $g_m(\eta)$ ($m\not= n$) and integrating the
resulting equation over 
$-\eta_{\rm s}$ to $\eta_{\rm s}$, we have 
\begin{eqnarray}
    -\int_{-\eta_{\rm s}}^{\eta_{\rm s}}{\rm exp}\biggr(-\frac{\eta^2}{2}\biggr)\biggr(\frac{dg_n}{d\eta}\frac{dg_m}{d\eta}-K_ng_ng_m\biggr)d\eta    \nonumber \\
    +\frac{2\omega_n^2}{\Omega_\bot^2}{\rm exp}\biggr(-\frac{\eta_{\rm s}^2}{2}\biggr)\frac{1}{\eta_{\rm s}}g_n(\eta_{\rm s})g_m(\eta_{\rm s})=0,
\label{orth2}
\end{eqnarray}
where integration by part has been applied by using the boundary condition (\ref{bc4}).
  
We have an equation similar to equation (\ref{orth2}) for $g_m$.
That is, starting from equation (\ref{orth11}) for $g_m$ and multiplying $g_n$ to both sides of the equation, we have a similar equation as equation (\ref{orth2}) after performing integration by part.
Then, taking the difference between the resulting equation and equation (\ref{orth2}), we have
\begin{eqnarray}
    (K_n-K_m)\int_{-\eta_{\rm s}}^{\eta_{\rm s}}{\rm exp}\biggr(-\frac{\eta^2}{2}\biggr)
                g_n(\eta)g_m(\eta) d\eta             \nonumber    \\
\hspace{30pt}      +\frac{2(\omega_n^2-\omega_m^2)}{\Omega_\bot^2}
                  {\rm exp}\biggr(-\frac{\eta_{\rm s}^2}{2}\biggr)\frac{1}{\eta_{\rm s}}g_n(\eta_{\rm s})
                  g_m(\eta_{\rm s}) =0.
\label{orth3}
\end{eqnarray}
The term resulting from surface integral in equation (\ref{orth3}) (i.e., the second term) does not 
vanish in general, since $\omega_n^2\not= \omega_m^2$.
Hence, there is no orthogonal relation such as equation (\ref{orth0}) in the present case of 
$\eta_{\rm s}\not= \infty$.
However, in the case of g-mode oscillations, the difference between $\omega_n^2$ and
$\omega_m^2$ is much smaller than $\Omega_\bot^2$ in the case of $\eta_{\rm s}\not=\infty$
(i.e., Okazaki et al. 1987).
Differences between $K_n$ and $K_m$ are on the order of unity.
Furthermore,  the surface value, ${\rm exp}(-\eta_{\rm s}^2/2)(1/\eta_{\rm s})g_n(\eta_{\rm s})
g_m(\eta_{\rm s})$, is smaller than the value of integration of the first term of equation (\ref{orth3}),
when $\eta_{\rm s}$ is not too small.
Hence, we have approximately  
\begin{equation}
         \int_{-\eta_{\rm s}}^{\eta_{\rm s}}{\rm exp}\biggr(-\frac{\eta^2}{2}\biggr)g_n(\eta)g(\eta)_m
                   d\eta \sim 0,
\label{orth1}
\end{equation}
when $n\not= m$.
We shall use this quasi-orthogonal relation in the following sections.

\section{Wave equation when $c_{\rm A}^2/c_{\rm s}^2$ is taken into account}

In the case of $c_{\rm A}^2/c_{\rm s}^2\not= 0$, 
the solution of equation (\ref{eqh1u}) does not have such a separable form as  
$h_1(r,\eta)=g(\eta)f(r)$.
If the effects of $c_{\rm A}^2/c_{\rm s}^2\not= 0$ on oscillations are weak,
however, the righthand side of equation (\ref{eqh1u}) can be treated as a small perturbation in
solving equation (\ref{eqh1u}).
That is, $h_1(r,\eta)$  can be approximately separated as $h_1(r,\eta)=g(\eta, r)f(r)$ with weak $r$-dependence of $g$.
This weak $r$-dependence of $g$ can be examined by a perturbation method.
It is noted here that in what cases the effects of $c_{\rm A}^2/c_{\rm s}^2$ can be treated as small perturbations can be found in the final results.

To proceed to this direction, we write
\begin{eqnarray}
    h_1(r,\eta)=f(r) g(\eta, r),      \nonumber \\
    u_r(r,\eta)=f_{r}(r)g_{r}(\eta,r),      \nonumber \\
    u_\varphi(r,\eta)=f_{\varphi}(r)g_{\varphi}(\eta,r),
\label{}
\end{eqnarray}
and divide equation (\ref{eqh1u}) by $fg$.
Then, the resulting equation can be approximately separated into two equations with a weakly
$r$-dependent separation constant $K(r)$ as
\begin{eqnarray}
&&      \frac{1}{g}\biggr(\frac{\partial^2}{\partial\eta^2}-\eta\frac{\partial}{\partial \eta}\biggr)g
              \nonumber    \\
&&      +\frac{1}{fg}H^2\biggr[\frac{\omega^2}{\omega^2-\kappa^2}\frac{df}{dr}\frac{\partial g}{\partial r}+
      \frac{\partial}{\partial r}\biggr(\frac{\omega^2}{\omega^2-\kappa^2}f\frac{\partial g}{\partial r}\biggr)  
             \biggr]
              \nonumber   \\
&&      +\frac{i\omega}{fg}\frac{\partial}{\partial r}\biggr[
      \frac{c_{\rm A}^2}{\omega^2-\kappa^2}\biggr(H^2\frac{\partial^2}{\partial r^2}(f_{r}g_{r})+
            f_{r}\frac{\partial^2 g_{r}}{\partial \eta^2}+\frac{2\Omega}{i\omega}
            f_{\varphi}\frac{\partial^2g_{\varphi}}{\partial \eta^2}\biggr)\biggr]
                \nonumber   \\
&&                  =-K(r)
\label{separation_g}
\end{eqnarray}
and
\begin{equation}
       \frac{\omega^2}{\Omega_\bot^2}+\frac{1}{f}H^2\frac{d}{dr}\biggr(\frac{\omega^2}{\omega^2-\kappa^2}\frac{df}{dr}\biggr) =K(r).
\label{separation_f}
\end{equation}

We now expand $g(\eta, r)$ as
\begin{equation}
         g(r,\eta)= g^{(0)}(\eta)+g^{(1)}(\eta,r)+....
\label{expansion}
\end{equation}
Here, $g^{(0)}$ is the zeroth-order solution obtained in the previous section with 
$c_{\rm A}^2/c_{\rm s}^2= 0$, and $g^{(1)}$ is the perturbed part of $g^{(0)}$ due to
$c_{\rm A}^2/c_{\rm s}^2\not= 0$.  
It is noted that $g(\eta)$ and $K$ in the previous section are hereafter denoted by attaching superscript
(0) as  $g^{(0)}(\eta)$ and $K^{(0)}$ 
in order to emphasize that they are the zeroth-order quantities. 

In the lowest order approximation where the terms of $c_{\rm A}^2/c_{\rm s}^2$ are neglected,
$g^{(0)}$ is independent of $r$ and equations (\ref{separation_g}) and (\ref{separation_f}) are reduced to
equations (\ref{eqg}) and (\ref{eqf}). 
Let us now proceed to the next order approximations where the terms of $c_{\rm A}^2/c_{\rm s}^2$
are considered.
Then, $g^{(1)}$ depends weakly on $r$, and also the separation constant also depends weakly
on $r$ as $K(r)=K^{(0)}+K^{(1)}(r)$.
Then, from equations (\ref{separation_g}) and (\ref{separation_f}), we have, respectively, 
\begin{eqnarray}
&&      \biggr(\frac{\partial^2}{\partial\eta^2}-\eta\frac{\partial}{\partial \eta}+K^{(0)}\biggr)g^{(1)}
              \nonumber    \\
&&      = -H^2\biggr[\frac{\omega^2}{\omega^2-\kappa^2}\frac{1}{f}\frac{df}{dr}\frac{\partial g^{(1)}}{\partial r}+
      \frac{\partial}{\partial r}\biggr(\frac{\omega^2}{\omega^2-\kappa^2}f\frac{\partial g^{(1)}}{\partial r} 
              \biggr)  \biggr]
              \nonumber   \\
&&      -\frac{i\omega}{f}\frac{\partial}{\partial r}\biggr[
      \frac{c_{\rm A}^2}{\omega^2-\kappa^2}\biggr(H^2\frac{d^2 f_r}{dr^2}g^{(0)}_{r}+
            f_{r}\frac{d^2 g_{r}^{(0)}}{d \eta^2}+\frac{2\Omega}{i\omega}
            f_{\varphi}\frac{d^2g_{\varphi}^{(0)}}{d \eta^2}\biggr)\biggr]
                \nonumber   \\
&&                  -K^{(1)}(r)g^{(0)}.
\label{separation_g1}
\end{eqnarray}
and
\begin{equation}
       H^2\frac{\partial}{\partial r}\biggr(\frac{\omega^2}{\omega^2-\kappa^2}\frac{df}{dr}\biggr) 
         +\frac{\omega^2}{\Omega_\bot^2}f =(K^{(0)}+K^{(1}) f.
\label{separation_f1}
\end{equation}
      
Equation (\ref{separation_g1}) is an inhomogeneous differential equation with respect to $g^{(1)}$,
and the lefthand side of equation (\ref{separation_g1}) is written in the same form as the 
zeroth-order equation (\ref{eqg}).
In the followings we solve equation (\ref{separation_g1}) by a standard perturbation method.
The standard perturbation method requires that the zeroth order eigenfunctions are orthogonal.
In the present problem we do not have exact orthogonality, but orthogonal relations are roughly
realized as shown in equation(\ref{orth1}).
We shall be satisfied by using this quasi-orthogonality, since the resulting errors seem to be small.

As in the standard perturbation method, 
$g^{(1)}$ is now expanded in terms of the set of eigenfunctions 
in the zeroth order equation (\ref{eqg}), say $g^{(0)}_m$, where $m=0$, 1, 2,... as
\begin{equation}  
             g^{(1)}(\eta, r)=\sum_{m=0}a_{m}(r)g_m^{(0)}(\eta)
\label{expansion_g1}
\end{equation}
and the coefficients $a_m(r)$'s ($m\not= 1$) are determined  by use of quasi-orthogonality of eigenfunctions $g_m^{(0)}$.
The perturbation method remains the coefficient $a_1(r)$ ($m=1$) undetermined, but
requires that the righthand side of (\ref{separation_g1}) is orthogonal to $g^{(0)}$.
This is the solvability condition of equation (\ref{separation_g1}).
It is noted that the coefficient $a_1(r)$ can be taken to be zero, since the term $a_1(r)g_1^{(0)}$ in expansion of
$g^{(1)}$ can be included in the zeroth order solution of $h_1=f(r)g^{(0)}$ (i.e., normalization of $f$)\footnote{
It is easily shown that if $a_1(r)$ is taken to be a non-zero arbitrary function of $r$,
the results become the same as the case where $h_1$ is written formally as $h_1=\tilde{f}(r)g$ with
$\tilde{f}=f(r)+a_1(r)$ and $a_1(r)$ in the expansion of $g^{(1)}(\eta, r)$ is taken to be zero.
In this case $\tilde{f}$ is found to follow the same equation as equation (\ref{separation_f1}).
}.

Using the quasi-orthogonality (\ref{orth1}), we see that the solvability condition that the righthand 
side of equation (\ref{separation_g1}) is orthogonal to $g^{(0)}$ is approximately written as 
\begin{equation}
            K^{(1)}f=
     -i\omega\frac{d}{dr}\biggr[\frac{c_{{\rm A}0}^2}{\omega^2-\kappa^2}
     \biggr\{AH^2\frac{d^2f_r}{dr^2}
    +B\biggr(f_r+\frac{2\Omega}{i\omega}f_{\varphi}\biggr)\biggr\}\biggr],
\label{K1}
\end{equation}            
where
\begin{eqnarray}
        A=  \int_{-\eta_{\rm s}}^{\eta_{\rm s}}g^{(0)}g^{(0)}d\eta
              \biggr/ 
             \int_{-\eta_{\rm s}}^{\eta_{\rm s}}{\rm exp}\biggr(-\frac{\eta^2}{2}\biggr)
                          g^{(0)}g^{(0)}d\eta,
\label{A}                          
\end{eqnarray}
\begin{eqnarray}
        B=  \int_{-\eta_{\rm s}}^{\eta_{\rm s}}g^{(0)}\frac{d^2g^{(0)}}{d\eta^2}d\eta
              \biggr/ 
             \int_{-\eta_{\rm s}}^{\eta_{\rm s}}{\rm exp}\biggr(-\frac{\eta^2}{2}\biggr)
                          g^{(0)}g^{(0)}d\eta.
\label{B}                          
\end{eqnarray}
In deriving solvability condition (\ref{K1}) we have used
\begin{equation}
        g_r^{(0)}(\eta)= g_{\varphi}^{(0)}(\eta)=g^{(0)}(\eta),
\end{equation}
which will be found later, and $c_{{\rm A}0}$ is the Alfv\'{e}n speed on the equator [see equation
(\ref{Alfven})]. 
The values of $A$ and $B$ are shown in table 1 for three cases of $\eta_{\rm s}=2.5$, $3.0$ and 5.0.
Values of $A$ and $B$ also depend weakly on $\omega^2/\Omega_\bot^2$.
Three cases of $\omega^2/\Omega_\bot^2=0.4$, 0.6, and 0.8 are shown for $\eta_{\rm s}=3.0$ and
5.0.
For comparison, values of $A$ and $B$ in the case of $\eta_{\rm s}=\infty$ are also shown.

\begin{table}
	\centering
	\caption{Values of $A$, $B$, and the conditions of $Ac_{{\rm A}0}^2/c_{\rm s}^2<1$
	   and of self-trapping for three cases of $\eta_{\rm s}$ with some values of 
	   $\omega^2/\Omega_\bot^2$.}
	\label{tab:example_table}	
	\begin{tabular}{lccc} 
	\hline
   $\eta_{\rm s}$ & $\omega^2/\Omega_\bot^2$ &$A$, $B$& condition of     \\
                  &          &     & $Ac_{{\rm A}0}^2/c_{\rm s}^2<1$ \\
   \hline   
        2.5  &  0.4   &  4.517,  -0.506            &  $c_{{\rm A}0}^2/c_{\rm s}^2< 0.22$  \\
  \hline          
	3.0  &   0.4 & 7.258, -0.549            & $c_{{\rm A}0}^2/c_{\rm s}^2< 0.14$ \\
	       &   0.6  &  7.305, -0.376 &      $c_{{\rm A}0}^2/c_{\rm s}^2< 0.14$  \\
	       &   0.8  &  7.356, -0.198 &      $c_{{\rm A}0}^2/c_{\rm s}^2< 0.14 $ \\
   \hline	
  	5.0  &   0.4 & 32.97,  -1.111  &       $c_{{\rm A}0}^2/c_{\rm s}^2< 0.030$ \\
	       &   0.6  & 33.09,  -0.743 &     $c_{{\rm A}0}^2/c_{\rm s}^2< 0.030$ \\
               &   0.8  &  33.20, -0.372 &     $c_{{\rm A}0}^2/c_{\rm s}^2< 0.030$ \\
 \hline	    
         $\infty$ &         &  $\infty$,   0.0   &   $c_{{\rm A}0}^2/c_{\rm s}^2= 0.0$ \\   
 \hline  
	\end{tabular}
\end{table}

To simplify equation (\ref{K1}) further, let us express $f_u$ and $f_\varphi$ by $f$.
In the limit of $c_{{\rm A}0}=0$, equation (\ref{h1ur2}) gives 
\begin{equation}
         (\omega^2-\kappa^2)u_r= i\omega\frac{\partial h_1}{\partial r}.
\label{ur}
\end{equation}
This shows that $g_r^{(0)}$ can be taken to be equal to $g^{(0)}$ and leads to
\begin{equation} 
        (\omega^2-\kappa^2)f_r= i\omega\frac{df}{dr}
\label{fu-f}        
\end{equation}
in the limit of $c_{\rm A}^2=0$.
Furthermore, in the limit of $c_{\rm A}^2=0$ the $\varphi$-component of equation of motion gives a relation between $f_r$ and $f_\varphi$, which is 
\begin{equation}
                f_\varphi=-\frac{i}{\omega}\frac{\kappa^2}{2\Omega}f_r=\frac{\kappa^2}{2\Omega}
                       \frac{1}{\omega^2-\kappa^2}\frac{df}{dr}.
\label{fvarphi-f}
\end{equation}

Substitution of equations (\ref{fu-f}) and (\ref{fvarphi-f}) into equation (\ref{K1}) 
gives an expression for $K^{(1)}f $ in terms of $f$ alone.
Substitution of this expression for $K^{(1)}f$ into equation (\ref{separation_f1}) leads finally to  
\begin{eqnarray}
       H^2\frac{d}{d r}\biggr[\frac{\omega^2}{\omega^2-\kappa^2}
               \biggr(1-B\frac{c_{{\rm A}0}^2}{c_{\rm s}^2}
                        \frac{\Omega_\bot^2}{\omega^2}\biggr)\frac{df}{dr}\biggr]
         +\frac{\omega^2-K^{(0)}\Omega_\bot^2}{\Omega_\bot^2}f 
               \nonumber    \\
     =H^4\frac{d}{dr}\biggr[A\frac{c_{{\rm A}0}^2}{c_{\rm s}^2}
        \frac{\Omega_\bot^2}
     {\omega^2-\kappa^2} \frac{d^2}{dr^2}\biggr(\frac{\omega^2}{\omega^2-\kappa^2}\frac{df}{dr}\biggr)\biggr].
\label{separation_f2}
\end{eqnarray}
This is a wave equation expressed in terms of $f$ alone.
The terms with $A$ and $B$ represent the effects of magnetic fields, 
which are taken into account as perturbations.
In treating this equation, however, we need careful considerations, since 
the order of derivative with respect to $r$ has been increased by the term with $A$
from that of the unperturbed one.
The unperturbed equation with no magnetic fields is 
\begin{equation}
     H^2\frac{d}{d r}\biggr[\frac{\omega^2}{\omega^2-\kappa^2}
               \frac{df}{dr}\biggr]
         +\frac{\omega^2-K^{(0)}\Omega_\bot^2}{\Omega_\bot^2}f =0.
\label{zeroth}
\end{equation}
The increase of the order of differential equation means that 
unless the terms resulting from the effects of $c_{{\rm A}0}^2/c_{\rm s}^2\not= 0$ are fully taken 
into account in equation (\ref{separation_f2}), there is the possibility that solutions of
(\ref{separation_f2}) do not tend to those of equation (\ref{zeroth}) in the limit of
$c_{{\rm A}0}^2/c_{\rm s}^2=0$.
This is due to the fact that the characteristics of differential equations are changed by
the change of order of equations.

To avoid this difficulties, we should remember that the fourth order term in 
equation (\ref{separation_f2}) is a term resulting from perturbations.
Hence, by substituting the zeroth order solution (\ref{zeroth}) into the righthand side
of equation (\ref{separation_f2}) we reduce the fourth order term 
to a second order term.
Then, we reduce equation (\ref{separation_f2}) to
\begin{eqnarray}
       H^2\frac{d}{d r}\biggr[\frac{\omega^2}{\omega^2-\kappa^2}
               \biggr(1-B\frac{c_{{\rm A}0}^2}{c_{\rm s}^2}
                        \frac{\Omega_\bot^2}{\omega^2}\biggr)\frac{df}{dr}\biggr]
         +\frac{\omega^2-K^{(0)}\Omega_\bot^2}{\Omega_\bot^2}f 
               \nonumber    \\
     =-H^2\frac{d}{dr}\biggr[A\frac{c_{{\rm A}0}^2}{c_{\rm s}^2}
        \frac{\Omega_\bot^2}{\omega^2-\kappa^2} 
        \frac{d}{dr}\biggr(\frac{\omega^2-K^{(0)}\Omega_\bot^2}{\Omega_\bot^2}f\biggr)
         \biggr].
\label{separation_f22}
\end{eqnarray}   
This equation is the final wave equation describing the axisymmetric g-mode oscillations. 

The righthand side of equation (\ref{separation_f22}) comes from the perturbation.
Thus, comparison of the righthand side of this equation with the first term 
on the lefthand side
roughly shows that equation (\ref{separation_f22}) is valid only when
\begin{equation}
        A\frac{c_{{\rm A}0}^2}{c_{\rm s}^2}\frac{\vert \omega^2-K^{(0)}\Omega_\bot^2\vert}{\omega^2}
          < 1. 
\label{condition-of-A}
\end{equation}
That is, we can study trapping of axisymmetric g-mode oscillations by use of equation 
(\ref{separation_f22}), as long as $c_{{\rm A}0}^2/c_{\rm s}^2$ is small in the sense 
that inequality (\ref{condition-of-A}) is satisfied.
Unless we consider very low-frequency oscillations, 
the above condition is roughly $Ac_{{\rm A}0}^2/c_{\rm s}^2<1$.
The values of $c_{{\rm A}0}^2/c_{\rm s}^2$ required for $Ac_{{\rm A}0}^2/c_{\rm s}^2<1$
are shown in table 1 for disks with some finite disk thickness 
(i.e., $\eta_{\rm s}\not=\infty$).

It is noted that the present perturbation method cannot be applied when 
$\eta_{\rm s}=\infty$ (see table 1), since in this case, $A=\infty$ and condition 
(\ref{condition-of-A}) requires $c_{{\rm A}0}^2/c_{\rm s}^2=0$.

\section{Propagation Region of G-Mode Oscillations and Self-Trapping}
 
Starting from equation (\ref{separation_f22}), we examine the propagation region of g-mode oscillations.
In equation (\ref{separation_f22}) there is an apparent singularity at the radius of
$\omega^2=\kappa^2$.
To avoid this inconvenience we introduce a new dependent function ${\tilde f}$ defined by
\begin{equation}
      {\tilde f}=\frac{\omega^2}{\omega^2-\kappa^2}\frac{df}{dr}.
\label{tild_f}
\end{equation}
This function ${\tilde f}$ is related to $u_r$ [see equation (\ref{fu-f})].
Furthermore, for simplicity, the radial variation of $(\omega^2-K^{(0)}\Omega_\bot^2)/\Omega_\bot^2$ 
in the wave propagation region is neglected.
The term with $B$ in equation (\ref{separation_f22}) is also neglected, because it is a small term.
Then, from equation (\ref{separation_f22}) we have
\begin{equation}
       \frac{d\tilde f}{dr}+\frac{\omega^2-K^{(0)}\Omega_\bot^2}{c_{\rm s}^2}f
           +A\frac{c_{{\rm A}0}^2}{c_{\rm s}^2}\frac{d}{dr}\biggr( 
                    \frac{\omega^2-K^{(0)}\Omega_\bot^2}{\omega^2}{\tilde f}\biggr)=0.
\label{tilde_f2}
\end{equation}
Taking the radial derivative of this equation, we have approximately
\begin{equation}
     \biggr(1+A\frac{c_{{\rm A}0}^2}{c_{\rm s}^2}
     \frac{\omega^2-K^{(0)}\Omega_\bot^2}{\omega^2}\biggr)
                           \frac{d^2{\tilde f}}{dr^2}
     +\frac{(\omega^2-\kappa^2)(\omega^2-K^{(0)}\Omega_\bot^2)}{c_{\rm s}^2\omega^2}{\tilde f}=0.
\label{tilde_f3}
\end{equation}

This equation tends to the wave equation describing oscillations in disks
with no magnetic fields in the limit of $c_{{\rm A}0}^2/c_{\rm s}^2=0$.
This equation clearly shows that the wave propagation region is specified by
$(\omega^2-\kappa^2)(\omega^2-K^{(0)}\Omega_\bot^2)>0$, when condition (\ref{condition-of-A})
is satisfied.
In the case of g-mode oscillations we have 
$\omega^2-K^{(0)}\Omega_\bot^2 <0$, and thus their
propagation region is $\omega^2-\kappa^2 < 0$.
In relativistic disks, the propagation region of g-mode oscillations with 
$\omega< \kappa_{\rm max}$ is bounded by two radii where $\omega^2=\kappa^2$ is realized,
where $\kappa_{\rm max}$ is the maximum value of epicyclic frequency.
Furthermore, at the radii where $\omega^2=\kappa^2$ is realized the wavenumber of
oscillations vanishes:
Waves approached there are reflected back.
That is, g-mode oscillations are trapped by cavity due to radial distribution of 
epicyclic frequency.
The above results show that the characteristics concerning the wave trapping region 
of g-mode oscillations in the case of no magnetic fields remains unchanged even in
magnetized disks, as long as condition (\ref{condition-of-A}) holds.

It will be instructive to show that the trapping is also derived from 
equation (\ref{separation_f2}) by taking $d/dr=-ik_r$.
That is, by taking $d/dr=-ik_r$ under the local approximation and other approximations adopted 
above, we have formally from
equation (\ref{separation_f2}):
\begin{equation}
      \frac{\omega^2}{\omega^2-\kappa^2}(kH)^2
      -\frac{\omega^2-K^{(0)}\Omega_\bot^2}{\Omega_\bot^2}
      +A\frac{c_{{\rm A}0}^2}{c_{\rm s}^2}\frac{\omega^2\Omega_\bot^2}
             {(\omega^2-\kappa^2)^2}(kH)^4=0.
\label{local2}
\end{equation}

This equation needs to be solved by regarding the last term as a perturbation term.
That is, $(kH)^2$ is written as $(kH)^2=(kH)_0^2+(kH)_1^2+...$.
Then, we have
\begin{equation}
      (kH)_0^2=\frac{(\omega^2-\kappa^2)(\omega^2-K^{(0)}\Omega_\bot^2)}
               {\omega^2\Omega_\bot^2}
\end{equation}
and 
\begin{equation}
       (kH)_1^2=-A\frac{c_{{\rm A}0}^2}{c_{\rm s}^2}
                 \frac{\Omega_\bot^2}{\omega^2-\kappa^2}(kH)_0^4.
\end{equation}
From the above two equations, we have
\begin{equation}
    (kH)^2=\frac{(\omega^2-\kappa^2)(\omega^2-K^{(0)}\Omega_\bot^2)}
               {\omega^2\Omega_\bot^2}
      \biggr[1-A\frac{c_{{\rm A}0}^2}{c_{\rm s}^2}
               \frac{\omega^2-K^{(0)}\Omega_\bot^2}{\omega^2}\biggr].
\end{equation}     
This expression for $(kH)^2$ is the same as that obtained from equation (\ref {tilde_f3})
by taking $d^2{\tilde f}/dr^2=-k^2{\tilde f}$, when condition (\ref {condition-of-A})
holdes.         

\section{Summary and Discussion} 

In non-magnetized relativistic disks, g-mode oscillations are self-trapped in the innermost
region of the disks (Okazaki et al. 1987).
This self-trapping of oscillations is of importance, because such discrete oscillations in
infinitely extended disks 
might be one of possible causes of high-frequency quasi-periodic oscillations observed 
in black-hole and neutron-star X-ray binaries. 
Among g-mode oscillations, however, non-axisymmetric ones are outside of our interest, 
because these oscillations have a corotation point in their propagation region except for special cases
and are strongly damped by the corotation resonance
(Kato 2003, Li et al. 2003).

Fu and Lai (2009), however, suggested that  all g-mode oscillations are strongly affected by the 
presence of poloidal magnetic fields and  
their self-trapping is easily destroyed even if the fields are weak.
Their results are of interest, but are based on a rough examination of wave motions 
in disks which extend infinitely (i.e., $\eta_{\rm s}=\infty$) in the vertical direction. 
Wave motions in vertically thick disks with vertical magnetic fields have complicated 
vertical behaviors.
The complications come from the fact that 
the kinetic energy density of waves,  for example $\rho_0u_r^2$, and that of magnetic energy density, for example $b_\varphi^2/4\pi$, have quite different $z$-dependences.
This complication is related to the fact that the acoustic speed, $c_{\rm s}$,  is assumed to have no $z$-dependence, while the Alfv\'{e}n speed, $c_{\rm A}$,  increases with $z$ and becomes infinite 
if the disk extends 
infinity in the vertical direction.

To avoid the above complication, it is helpful to consider disks with finite vertical thickness.
Consideration of such disks is, however, not temporizing.
This is important and necessary.
Observations suggest that quasi-periodic oscillations occur  in a disk-corona system where  
disks are truncated at a certain height and sandwiched by hot coronae (Remillard 2005). 
Hence, we have considered in this paper the disks which are truncated at a certain height
by a hot corona.

The results of our analyses show that as long as dimensionless parameter, $Ac_{{\rm A}0}^2/c_{\rm s}^2$,
is smaller than unity (more rigorously, as long as the condition (\ref{condition-of-A}) 
is satisfied), axisymmetric g-mode oscillations are trapped in the cavity of
epicyclic frequency.
The magnetic fields required for $Ac_{{\rm A}0}^2/c_{\rm s}^2<1$ is small, but will not be
too small compared with those in realistic objects.
It should be noted that as vertical thickness of disks decreases, the critical value of
$c_{{\rm A}0}^2/c_{\rm s}^2$ for which our perturbation method is applicable increases
(see table 1).
In the limit of no termination of disk thickness (i.e., $\eta_{\rm s}=\infty$), however,
our perturbation method is not applicable, i.e., the required value of 
$c_{{\rm A}0}^2/c_{\rm s}^2$ tends to zero.
That is, in vertically extended disks characteristics of trapping of g-mode oscillations
are strongly affected even if the magnetic fields are weak.
Our analyses cannot say how the trapping is affected in such disks with strong vertical 
magnetic fields.

Quasi-periodic oscillations are not always observed in every black-hole and neutron-star sources,
and also their frequencies are not always robust except for those in some black-hole sources. 
High frequency QPOs may not be a homogeneous class.
Hence, trapped g-mode oscillations will still remain as one of possible candidates of QPOs
at least in sources with weak vertical magnetic fields.

An issue to be mentioned here is whether axisymmetric g-mode oscillations can be really excited in 
disks, since they will be damped by viscous processes unlike the p-mode oscillations
(see, e.g., Kato 1978, 2016).
We think that the most promising excitation process of axisymmetric g-mode oscillations is 
stochastic one by turbulence, although
any numerical simulations do not suggest the presence of this possibility in disks yet.
Non-radial oscillations in the Sun and stars are believed now to be excited by stochastic  processes
by turbulence (Goldreich and Keeley 1977, see also Kato 1966).
Similar processes are also naturally expected in disks, since the turbulence in disks are much 
stronger than in the Sun and stars (see Nowak and Wagoner 1993 and Kato 2016).
Another possible excitation process of axisymmetric g-mode oscillations
is a wave-wave resonant excitation process in deformed disks (Kato 2013).
In this case another oscillation which becomes a set of the g-mode oscillation is excited.
A question concerning this wave-wave resonant excitation is whether a disk deformation 
required for the resonance process is really expected in disks.

\bigskip
The author thanks the referee for invaluable suggestions, which gave  the author
an opportunity to much improve the original manuscript.

\end{document}